\documentclass{WileyMSP-template}
\usepackage{amsmath}
\usepackage{xcolor}
\usepackage{enumitem}
\newcommand{\BLG}{bilayer graphene}

\begin{document}

\pagestyle{fancy}
\rhead{\includegraphics[width=2.5cm]{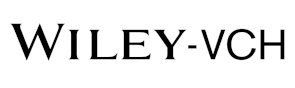}}

\title{Tuning confined states and  valley g-factors by quantum dot design in bilayer graphene}

\maketitle


\author{Dennis Mayer*}
\author{Angelika Knothe}



\begin{affiliations}
D. Mayer, Dr. A. Knothe\\
Address: Institut für Theoretische Physik, Universität Regensburg\\
Universitätsstraße 31, D-93053 Regensburg, Germany\\
Email Address: Dennis.Mayer@stud.uni-regensburg.de\\
Angelika.Knothe@physik.uni-regensburg.de

\end{affiliations}



\begin{abstract}

Electrostatically confined quantum dots in bilayer graphene have shown potential as building blocks for quantum technologies. To operate the dots, e.g., as qubits, a precise understanding and control of the confined states and their properties is required. Here, we perform large-scale numerical characterization of confined quantum states in bilayer graphene dots over an extensive range of gate-tunable parameters such as the dot size, depth, shape, and the bilayer graphene gap. We establish the dot states’ orbital degeneracy, wave function distribution, and valley g-factor and provide the parametric dependencies to achieve different regimes. We find that the dot states are highly susceptible to gate-dependent confinement and material parameters, enabling efficient tuning of confined states and valley g-factor modulation by quantum dot design.

\end{abstract}

\section{Introduction}
There has been a long-standing interest in graphene-based nanostructures due to carbon's low spin-orbit and hyperfine coupling promising long coherence times \cite{trauzettelSpinQubitsGraphene2007, fischerHyperfineInteractionElectronspin2009, falkoQuantumInformationChicken2007}. Recently, advanced fabrication and gating techniques have enabled the experimental realisation of electrostatically confined nanostructures in bilayer graphene, i.e., quantum channels \cite{overwegElectrostaticallyInducedQuantum2018, overwegTopologicallyNontrivialValley2018, leeTunableValleySplitting2020, goldCoherentJettingGateDefined2021, ingla-aynesSpecularElectronFocusing2023},  dots \cite{eichSpinValleyStates2018, eichCoupledQuantumDots2018, garreisCountingStatisticsSingle2023,  banszerusSpinpolarizedCurrentsBilayer2019, banszerusGateDefinedElectronHole2018, mollerProbingTwoElectronMultiplets2021, tongTunableValleySplitting2021, garreisShellFillingTrigonal2021,  arXiv:2305.09284, kurzmannKondoEffectSpin2021, gachterSingleShotSpinReadout2022, banszerusDispersiveSensingCharge2020, banszerusElectronHoleCrossover2020, arXiv:2303.10201, banszerusPulsedgateSpectroscopySingleelectron2020, banszerusSpinRelaxationSingleelectron2022, banszerusSpinvalleyCouplingSingleelectron2021, 
geVisualizationManipulationBilayer2020, velascoVisualizationControlSingleElectron2018, PhysRevLett.128.206805, PhysRevLett.124.166801}, 
Josephson junctions, interferometers, and SQUIDS \cite{iwakiriGateDefinedElectronInterferometer2022, rodan-legrainHighlyTunableJunctions2021, devriesGatedefinedJosephsonJunctions2021, portolesTunableMonolithicSQUID2022, dauberExploitingAharonovBohmOscillations2021, elahiDirectEvidenceKleinantiKlein2022}, in an endeavor to design bilayer graphene-based nanostructures for future quantum technology applications.

\begin{figure}[h]
  \includegraphics[width=\linewidth]{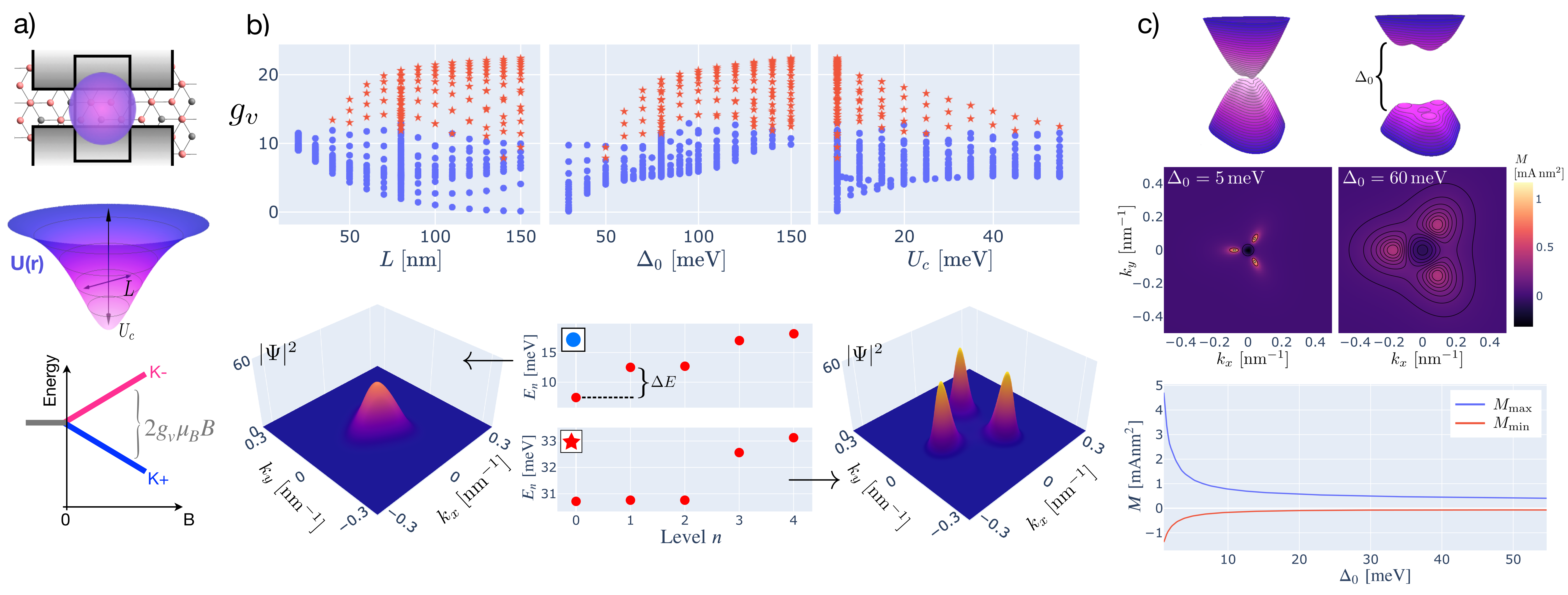}
  \caption{a) A combination of gates (including split gates and a finger gate)  electrostatically confines charge carriers in bilayer graphene into a quantum dot (top). The gate-induced confinement potential, $U(\mathbf{r})$, is smooth and gate-tunable (center). The dominant source of state splitting in a weak perpendicular magnetic field is valley splitting characterized by the valley g-factor, $g_v$ (bottom). b) Scatterplots visualizing the statistical correlations between the lowest-orbital dot state's $g_v$ and the dot parameters $L, U_c$, and the bilayer graphene gap $\Delta_0$. We distinguish the \textit{valley center regime} (when the lowest energy orbital is non-degenerate and the dot wavefunction is peaked within the center of the valley in momentum space, marked by blue dots) and the \textit{minivalley regime} (lowest orbital threefold degenerate with wavefunctions in the minivalleys, marked by red stars). The examples in the bottom panel were computed using the parameters $L$=20 nm, $\Delta_0$=60 meV, $U_c$=20 meV (valley center regime) and $L$=80 nm, $\Delta_0$=150 meV, $U_c$=20 meV (minivalley regime), respectively. c) Electronic and topological properties of the bilayer graphene Bloch bands as a function of the gap. The bands are almost quadratic around the Brillouin zone corners (valleys at the K-points) for small gaps, and the Berry-curvature-induced orbital magnetic moment, $M$, exhibits strong peaks and dips. At large gaps, three minivalleys develop around each K-point, while the maximum and minimum value of M is suppressed. Plots for the $K^{-}$ valley where the $K^{+}$ valley is rotated by $\pi$ with the opposite sign of $M$. }
  \label{fig:1}
\end{figure}

In these electrostatically defined structures, one uses a combination of multiple gates (including a split gate and a crossing finger gate, c.f. \textbf{Figure \ref{fig:1} a)}) to locally modulate the bilayer graphene gap and the charge carrier density in order to confine individual electrons. In such setups, the bilayer graphene band structure, as well as the confinement potential, are simultaneously tuned by the gates. This gate-tunability enables immense control and the opportunity to manipulate the confined quantum states. Therefore, a thorough understanding is required of how the confined states depend on the gate-tunable parameters and their interplay in order to design tailor-made bilayer graphene quantum nanostructures, e.g., quantum dots to be operated as qubits.

We present a large-scale space exploration investigating the dependence of confined bilayer graphene dot states on the gate-tunable dot and material parameters. Previous studies of bilayer graphene quantum dots have provided proof-of-principle calculations \cite{knotheQuartetStatesTwoelectron2020, knotheTunnelingTheoryBilayer2022} or descriptions of single points in parameters space to characterize specific experiments \cite{garreisShellFillingTrigonal2021, tongTunableValleySplitting2021, arXiv:2305.09284, mollerProbingTwoElectronMultiplets2021}. Here, in turn, we offer the full parameter dependence, giving tangible indications of how to design the confinement in order to tailor the dot states and their properties to achieve different regimes and functionalities. By numerically investigating bilayer graphene quantum dots over an extensive range of differently shaped soft confinements, we study:
\begin{itemize}[align=right,itemindent=2em,labelsep=2pt,labelwidth=1em,leftmargin=0pt,nosep]
    \item  The confined states in circular dots, their orbital degeneracy, and the corresponding wave functions. We find a regime where the orbital ground state is singly degenerate, and the wave function is localized around the center of the valley in momentum space, compared to a regime where the ground state acquires a threefold degeneracy corresponding to three minivalleys around the valley center. We give the parametric dependencies on the dot size and depth and the bilayer graphene gap for achieving either of the regimes.
\item The confined states' valley g-factor, which descends from the bilayer graphene Bloch bands' Berry curvature and determines the valley splitting of the confined states in a weak perpendicular magnetic field. We establish the dependencies of the valley g-factor on the dot parameters in either of the before-mentioned regimes and thereby demonstrate how to tune the valley splitting by quantum dot design, e.g., how to maximize or minimize the valley g-factor. 
    \item  How to modify the dot states and their properties via alterations of the dot shape. We find that elliptical deformations of the confinement potential allow one to tune further the wave function distribution, the orbital degeneracy, and the valley g-factor of the confined states.
\end{itemize}
For the dependencies of the dot states' characteristics on the confinement and material properties, we perform statistical analysis to reveal the correlations and provide a physical picture in terms of the underlying confined quantum states. The insights above will be relevant for efficient quantum dot design, e.g., for identifying suitable two-level regimes to operate the dots as qubits. Being able to modulate the g-factor by quantum dot design is also of interest for multi-dot setups with varying valley splitting across the different dots, allowing, e.g., for valley valves and selective state addressing.

\section{Microscopic modelling of confined dot states}

We describe the gate-defined soft confinement by the spatially varying confinement potential, $ U(\mathbf{r})$, sketched in Figure \ref{fig:1} a), and band gap profile, $\Delta(\mathbf{r})$,
\begin{align}
    U(\mathbf{r}) = U_c /  &\mathrm{cosh} \left( \frac{\sqrt{(\frac{x}{a})^2 + (\frac{y}{b})^2}}{L} \right),\, \mathrm{and} \\
    \Delta(\mathbf{r}) = \Delta_{0} -  \Delta_{mod} / &\mathrm{cosh} \left( \frac{\sqrt{(\frac{x}{a})^2 + (\frac{y}{b})^2}}{L} \right),
    \label{eqn:Potentials}
\end{align}
where the gate- and fabrication-defined parameters such as the dot depth, $U_c$, width $L$, and ellipticity $\frac{a}{b}$, and the pristine \BLG{} gap away from the dot can be varied independently. Further,  $\Delta_{mod}=0.3\Delta_0$ describes the gate-modulation of the gap towards the center of the dot. This specific value of the gate-induced gap modulation has been chosen in accordance with recent experiments, and we show the stability of our results with respect to slight variations in the appendix. 

Spatially varying potentials of the form as in Equation \eqref{eqn:Potentials} have proven successful in providing faithful, realistic models of gate-defined quantum dots in bilayer graphene in comparison with experiments \cite{knotheQuartetStatesTwoelectron2020, garreisShellFillingTrigonal2021, tongTunableValleySplitting2021, arXiv:2305.09284}. The potentials enter into the four-band Hamiltonian of \BLG{} \cite{mccannLowEnergyElectronic2007, mccannElectronicPropertiesBilayer2013}, 
\begin{align}
 \text{H}_{\xi}\!=\! \!&
\setlength{\arraycolsep}{+5pt} \begin{pmatrix} 
 U -\xi\frac{1}{2}\Delta  & \xi v_3\pi & 0 &\xi v \pi^{\dagger}\\
\xi v_3 \pi^{\dagger}&  U +\xi\frac{1}{2}\Delta  & \xi v\pi &0\\
 0 &  \xi v\pi^{\dagger} &   U +\xi\frac{1}{2}\Delta  &   \gamma_1\\
\xi v\pi & 0 &   \gamma_1 &  U -\xi\frac{1}{2}\Delta 
\end{pmatrix},
\label{eqn:H}
\end{align}
with momenta $\pi=p_x+ip_y,\,  \pi^{\dagger}=p_x-ip_y$, and parameters $v=1.02 \cdot 10^6 \text{ m/s}$,  $v_3\approx0.12 v$, and  $\gamma_1\approx0.38\text{ eV}$. Equation \eqref{eqn:H} is written for the two valleys $K^{\xi}$ labeled by the valley index $\xi=\pm 1$ and  in the Bloch basis $\psi_{K^+}=(\psi_{A},\psi_{B^{\prime}},\psi_{A^{\prime}},\psi_{B})$ in valley $K^+$, and $\psi_{K^-}=(\psi_{B^{\prime}},\psi_{A},\psi_{B},\psi_{A^{\prime}})$ in valley $K^-$  (in terms of the electron's amplitudes on the \BLG{} sublattices  $A$ and $B$ in the top, and  $A^{\prime}$ and $B^{\prime}$ in the bottom layer).

In order to compute the confined dot states, we employ an established numerical technique described in References \cite{knotheInfluenceMinivalleysBerry2018, knotheQuartetStatesTwoelectron2020, overwegTopologicallyNontrivialValley2018, leeTunableValleySplitting2020, garreisShellFillingTrigonal2021, tongPauliBlockadeTunable2022, arXiv:2305.09284}, where the Hamiltonian in Equation \eqref{eqn:H} is diagonalized in a suitable basis of localized states. From this diagonalization, we obtain the quantum dot spectrum, i.e., the eigenenergies of the $n$-th orbital dot level, $E_{\xi,n}$, and the corresponding eigenstates, $\Psi_{\xi,n}$. In the absence of any symmetry breaking, each orbital state is four-fold degenerate in the spin- and valley degree of freedom.

One key property of these quantized dot states is their coupling to an external magnetic field. Over the last few years, it has been shown that the dominant source of state splitting in a perpendicular magnetic field is the splitting of the two valleys $K^+$ and $K^-$. The corresponding valley g-factor, $g_v$, quantifying this valley splitting in weak magnetic fields, can be orders of magnitude larger than the spin g-factor \cite{eichSpinValleyStates2018, leeTunableValleySplitting2020, knotheQuartetStatesTwoelectron2020, tongTunableValleySplitting2021, arXiv:2305.09284}. The origin of this large valley splitting is the topological properties of bilayer graphene. Its Bloch bands, $|\alpha\rangle$, feature non-trivial Berry curvature, 
\begin{equation}
  \Omega_{\xi,\alpha} = -2 \hbar^2 \text{Im}\sum_{\alpha\neq \beta} \frac{\langle \alpha | \partial_{p_x} H_{\xi}|\beta\rangle\langle \beta | \partial_{p_y}H_{\xi} |\alpha\rangle}{(\varepsilon_{\alpha}-\varepsilon_{\beta})^2},
\end{equation}
entailing an orbital magnetic moment,  \cite{parkValleyFilteringDue2017, Fuchs2010, xiaoBerryPhaseEffects2010, changBerryPhaseHyperorbits1996, moulsdaleEngineeringTopologicalMagnetic2020}, 
\begin{equation}
    M_{\xi,\alpha} = - e \hbar \text{Im}\sum_{\alpha\neq \beta} \frac{\langle \alpha | \partial_{p_x} H_{\xi}|\beta\rangle\langle \beta | \partial_{p_y}H_{\xi} |\alpha\rangle}{\varepsilon_{\alpha}-\varepsilon_{\beta}},
    \label{eqn:M}
\end{equation}
where, for a band $\alpha$, we sum over the other three bands $\beta\neq\alpha$ in Equation \eqref{eqn:H} (For the confined states in the following, we consider states in the first conduction band,  dropping the index $\alpha$). The orbital magnetic moment, $M_{\xi}$, has opposite signs in the two valleys $K^+$ and $K^-$ and couples linearly to weak perpendicular magnetic fields, leading to sizeable magnetic field valley splittings. 

Since lowest orbital confined states are zero angular momentum states having both radial (principal) and azimuthal (angular momentum) quantum numbers close to zero (akin to the Fock-Darwin levels \cite{fockBemerkungZurQuantelung1928, darwinDiamagnetismFreeElectron1931, knotheQuartetStatesTwoelectron2020}), for these states the orbital magnetic moment of Equation \eqref{eqn:M} is the dominant quantity coupling to a perpendicular magnetic field. We can hence quantify the valley splitting of the lowest orbital confined dot states by integrating over the orbital magnetic moment and the wave function distribution in momentum space, \cite{leeTunableValleySplitting2020, tongTunableValleySplitting2021, knotheQuartetStatesTwoelectron2020, arXiv:2305.09284}
\begin{equation}
-\xi{g}_v=\frac{1}{\mu_B}\int d\boldsymbol{k} M_{\xi}(\boldsymbol{k} )|\Psi_{\xi,n=0}(\boldsymbol{k} )|^2,
\label{eqn:gv}
\end{equation}
determining how much the states in either valley are pushed up and down in a weak magnetic field. Equation \eqref{eqn:gv} demonstrates the subtle interplay of confinement and material properties as it depends on both the distribution of the confined dot wave functions and bilayer graphenes' orbital magnetic moment in momentum space.

As the properties of the dot states depend so intricately on the material characteristics of bilayer graphene as the host material, we demonstrate the electronic and topological properties of pristine bilayer graphene in Figure \ref{fig:1} c) as described by Hamiltonian in Equation \eqref{eqn:H} in the absence of any confinement potential and gap modulation, i.e., for $U\equiv 0$ and $\Delta\equiv \Delta_0$. Of the four resulting bands, the two low-energy bands are trigonally warped, forming three minivalleys around each of the K-points. The depth and the separation of these minivalleys increase with an increasing gap, such that the low-energy bands range from almost quadratic at small gaps to strongly modulated at large gaps. The orbital magnetic moment M inherits the trigonally warped triplet structure, with three peaks around the minivalleys. Due to the inverted, hole-like part of the band around the K-point, M exhibits a region of inverted sign around this area of momentum space. Both the maximum value of both the peaks and the dips in M increase as the bilayer graphene band gap decreases.


\section{Tuning dot states and the valley g-factor}

\subsection{States in a circular dot}

\begin{figure}[htbp]
  \includegraphics[width=\linewidth]{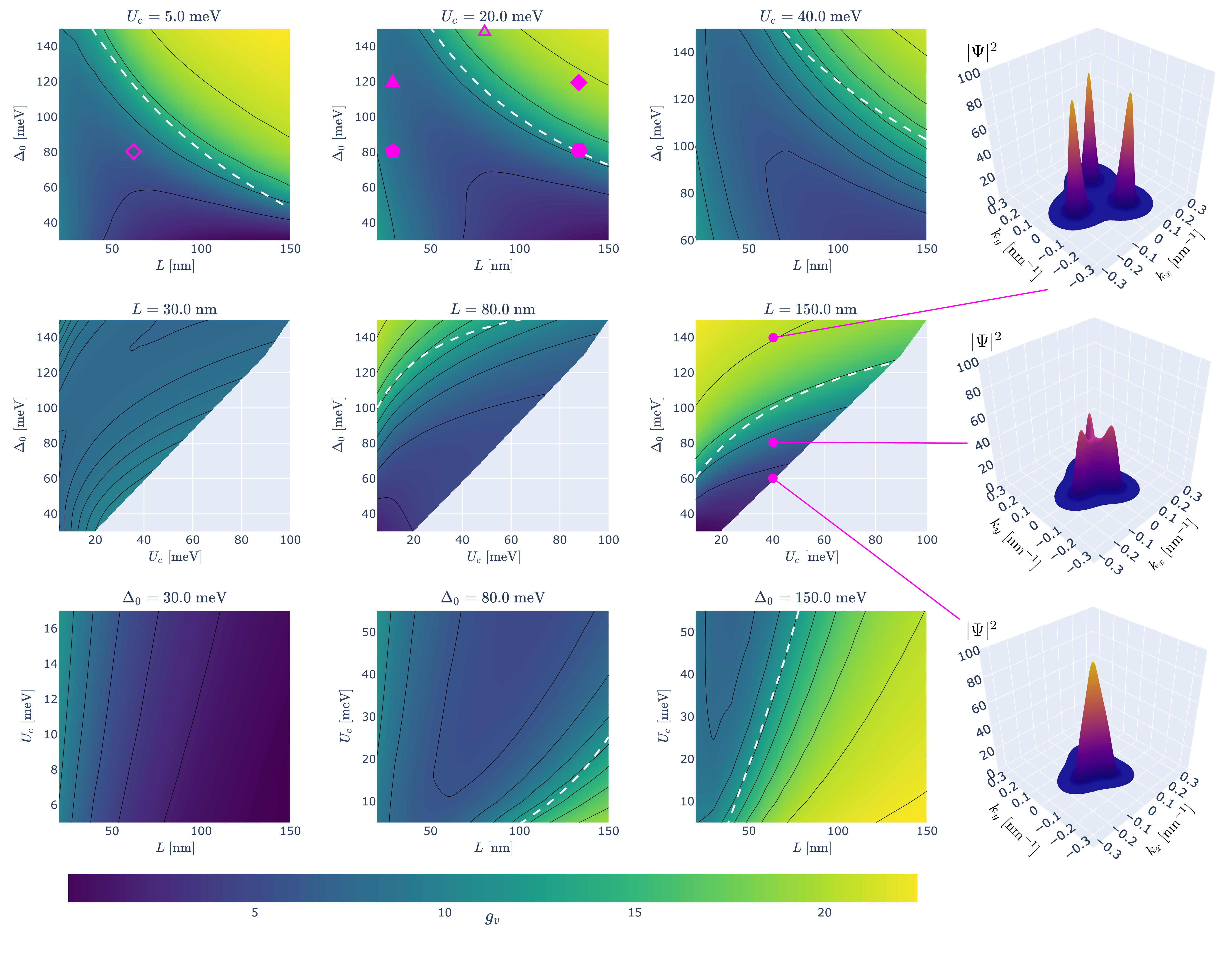}
  \caption{Evolution of the lowest orbital dot states' valley g-factor, $g_v$, over an extensive range of system parameters, $U_c, L,$ and $\Delta_0$. The dashed white lines indicate the transition from the valley-center regime (regions of low $g_v$, blue) to the minivalley regime (regions of high $g_v$, yellow). On the right, we show examples of the wave function distribution in momentum space characterizing these regimes (from minivalleys on the top to valley center on the bottom). We observe non-monotonic dependencies of $g_v$ on the gap size, width, and gap, depending on the wave functions momentum space distribution in the respective regimes. The filled pink symbols illustrate the positions in the parameter space for the examples in Figure \ref{fig:3} while the empty symbols relate to the dot states we elliptically deform in Figure \ref{fig:4}—all results for the $K^{-}$ valley.}
  \label{fig:2}
\end{figure}

We analyze the dot's ground state and their valley g-factor over an extensive range of gate-tunable parameters $\Delta_0, U_c$, and $L$ in a circular dot. Figure \ref{fig:1} b) depicts the statistical correlations between $g_v$ and the gap, the dot depth, and the width. These correlations illustrate what parametric regimes are most suitable when aiming at, e.g., minimizing or maximizing the value of $g_v$. Furthermore, we visualize the correlation between smaller/larger values of the valley g-factor and distinct momentum space distribution patterns of the confined dot wave function. The confined state's distribution ranges between being peaked around the K-point (the "valley-center" regime, which we characterize by non-zero weight at $K$ and a finite separation of the lowest and the first excited single-particle orbital, $\Delta E$) and sitting mainly in the three minivalleys around the K-point, with zero weight at the valley center (the "mini-valley" regime). In the latter case, the lowest three orbitals corresponding to the three minivalleys are degenerate.

In \textbf{Figure \ref{fig:2}}, we further characterize these regimes by analyzing the wave function's evolution and the valley g-factor across characteristic system parameter ranges. An overarching trend emerges, where the valley g-factor is consistently maximal within the minivalley regime. However, the dependence of $g_v$ on the system parameters is not always monotonic. 

From Figures 1 and 2, we deduce that the most suitable parameters for the valley center regime are small and deep dots, while large and shallow dots can promote the minivalley regime for sufficiently large values of the bilayer graphene gap.

\begin{figure}[t!]
  \includegraphics[width=1\linewidth]{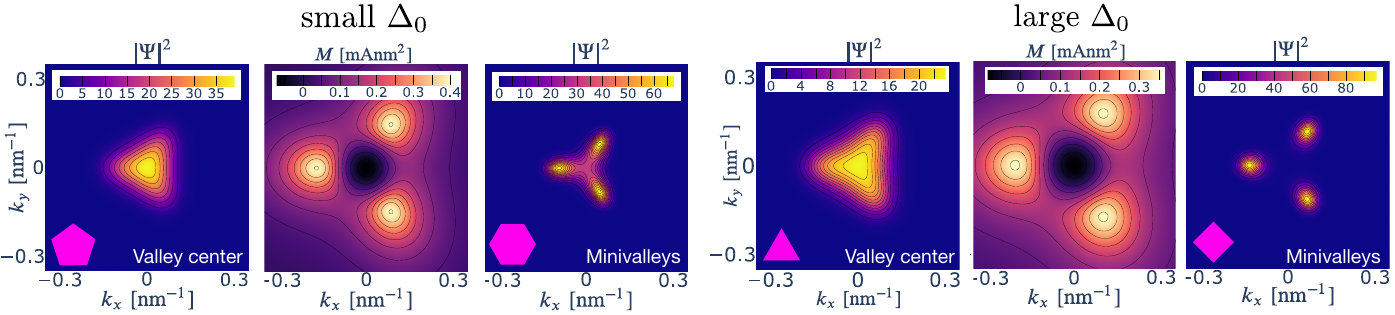}
  \caption{Gap dependence of the orbital magnetic moment and the lowest orbital confined dot wave functions in the valley center regime ($L$=30nm, $U_c$=20meV) compared to the minivalley regime ($L$=130nm, $U_c$=20meV) exemplified for a small gap (left, $\Delta_0$=80meV), and a large gap (right, $\Delta_0$=120meV) in valley $K^{-}$. Dot wave functions confined in the center of the valley are sensitive to the inverted part of the bilayer graphene band structure and hence to the region of negative M. Wave functions residing in the minivalleys pick up close to the maximal possible value of M near the band edges. The pink symbols relate to the symbols in Figure \ref{fig:2}.}
  \label{fig:3}
\end{figure}

Within these two regimes, the valley g-factor shows distinct dependence on the gap:  For fixed $L$ and $U_c$, the valley g-factor can decrease with increasing $\Delta_0$ within the valley center regime, while it generally increases strongly with an increasing gap in the minivalley regime (cf. the evolution between the filled pink symbols in the top center panel of Figure \ref{fig:2}). As we demonstrate in \textbf{Figure \ref{fig:3}}, this non-monotonic dependence of $g_v$, in particular on $\Delta_0$, stems from the influence of the hole-like part of the dispersion around the valley center where $M$ becomes negative. For small gaps, both the maximum peaks of M around the minivalleys and the minimum dip in the center increase in magnitude (cf. Figure \ref{fig:1} c)), entailing a partial cancellation when calculating $g_v$ for a wave function in the valley center regime. Conversely, within the minivalley regime, increasing the bilayer graphene gap promotes the minivalleys and further pushes the wave function towards the band edges, where M is peaked. Since, in this regime, the wave function has no weight around the center of the valley, the corresponding confined states are obnoxious to any negative values of the orbital magnetic moment. 

As a consequence, while the maximal value of the orbital magnetic moment generally is a decreasing function of the band gap (Figure \ref{fig:1} c)), the dependence of a dot state's $g_v$ is more subtle and depends on the distribution of the dot wave function in momentum space. The valley g-factor of the confined states in the mini-valley regime is closest to the maximal value achievable from the band structure's orbital magnetic moment.

The above discussion demonstrates how one may leverage the different dot parameters to tune the confined dot wave function, and hence the state orbital multiplicity and their magnetic field coupling via valley g-factor.

\subsection{States in elliptical dots}

One additional way to tune the wave function distribution and its symmetry is to alter the shape of the confinement potential compared to a circular quantum dot. Here, we study elliptical deformations since elliptical dot shapes are ubiquitous in experimental studies and representative of breaking the rotational symmetry along the two crystallographic directions of bilayer graphene.

\begin{figure}[t]
  \includegraphics[width=\linewidth]{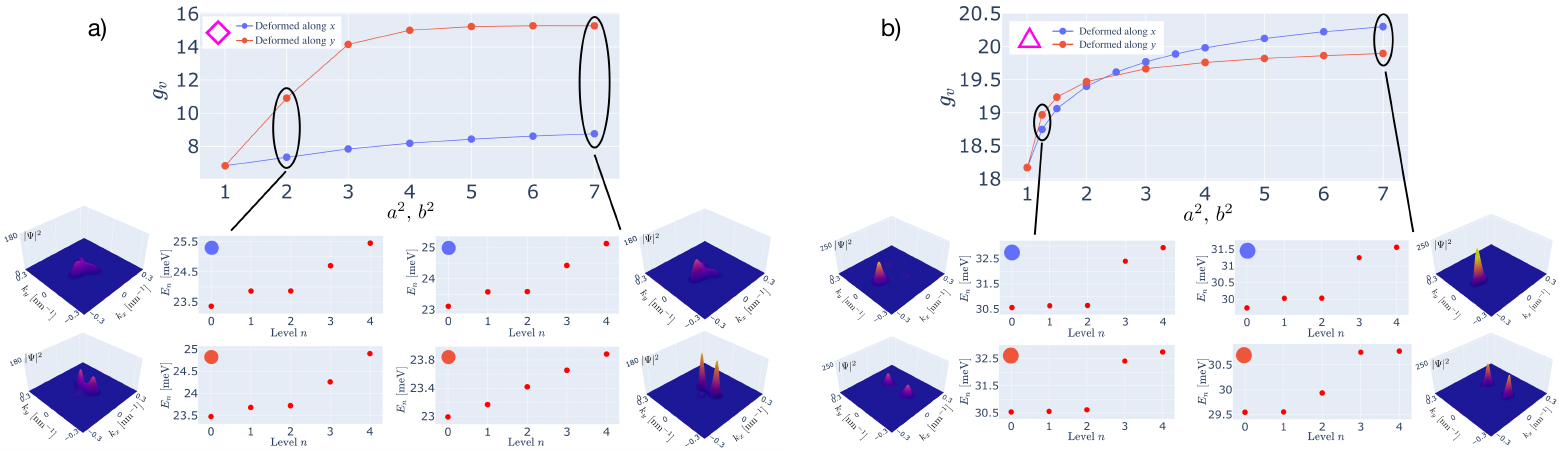}
  \caption{Top: dependence of the lowest orbital confined state's valley g-factor, $g_v$, on the elliptic deformation of the confinement potential. Here, the deformation parameters $a^2$ and $b^2$ quantify the elongation along the $x$- and $y$-axis, respectively, as prescribed by Equation \eqref{eqn:Potentials}.   The parameters are chosen such that states of the circular dot are a) in the valley center regime ($L$=60 nm, $\Delta_0$=80 meV, $U_c$=5 meV), and b) in the minivalley regime ($L$=80 nm, $\Delta_0$=150 meV, $U_c$=20 meV), respectively. The empty symbols relate to the symbols shown in Figure \ref{fig:2}. Bottom: Examples for the corresponding dot spectra and the lowest orbital confined state. All results for the $K^{-}$ valley.}
  \label{fig:4}
\end{figure}

\textbf{Figure \ref{fig:4}} exemplifies how the confined states and their valley g-factors evolve with the ellipticity of the dot for two characteristic parameter choices for the valley center regime and the minivalley regime, respectively. Overall, we observe an increase in the valley g-factor when the dot is deformed into an ellipse. This increase occurs as the dot wave function is pushed either into one of the minivalleys when the dot is deformed along the x-axis or into two minivalleys for deformations along the y-axis. In addition, breaking the threefold rotational symmetry between the three minivalleys also affects the orbital degeneracy of the ground state: States peaked in one minivalley are orbitally non-degenerate,  while states peaked in two minivalleys exhibit a two-fold orbital degeneracy.

We observe that the states in wide dots and for large gaps are most easily deformed, while deep confinement makes deforming the states more difficult. As a consequence, the states in the original minivalley regime, Figure \ref{fig:4} b), are most susceptible to being deformed in momentum space. The states in the original valley center regime, Figure \ref{fig:4} a), are more stable against deformations. The relative deformation-induced increase of the valley g-factor, however, is larger in the valley center regime as compared to the original minivalley regime since states with little or no support in the center of the valley do not suffer from cancellations due to the negative regions of the orbital magnetic moment, $M$, cf. Figure \ref{fig:3}.

The analysis above shows that realizing strongly elliptical bilayer graphene quantum dots represents one possible route to prepare confined states with large valley splittings at moderate gaps and dot sizes compared to the large dots and gaps required for the full minivalley regime in circular dots, cf. Figure \ref{fig:2}.

 \section{Conclusion}

 Overall, we have presented the confined bilayer graphene dot states, their spectra, and valley g-factors over an extensive range of gate-tunable parameters, such as the confinement, width, gap, shape, and the bilayer graphene gap. We find that the ground state can range from being non-degenerate (and peaked in the center of the valley in momentum space) to threefold degenerate (and sitting in the minivalleys) in circular dots and peaked in one or two minivalleys (with corresponding single or two-fold degeneracy) in elliptically deformed dots. Alongside the distribution of the confined dot wave functions, the corresponding topological valley g-factor changes in the different regimes as a function of the dot parameters and shape. The largest valley g-factors are observed when the wave function is peaked in the minivalleys, both in circular and elliptical dots. We note that we expect many of our observations, as long as they are based on wave function distributions, to generalize to higher dot levels with more complex wave function shapes. However, for higher confined dot states with non-zero angular momentum quantum numbers, the intrinsic angular magnetic moment needs to be considered in conjunction with the topological orbital magnetic moment when studying the coupling of the states to a magnetic field and the corresponding valley g-factors \cite{knotheQuartetStatesTwoelectron2020}. 

 \begin{figure}[h!]
  \includegraphics[width=\linewidth]{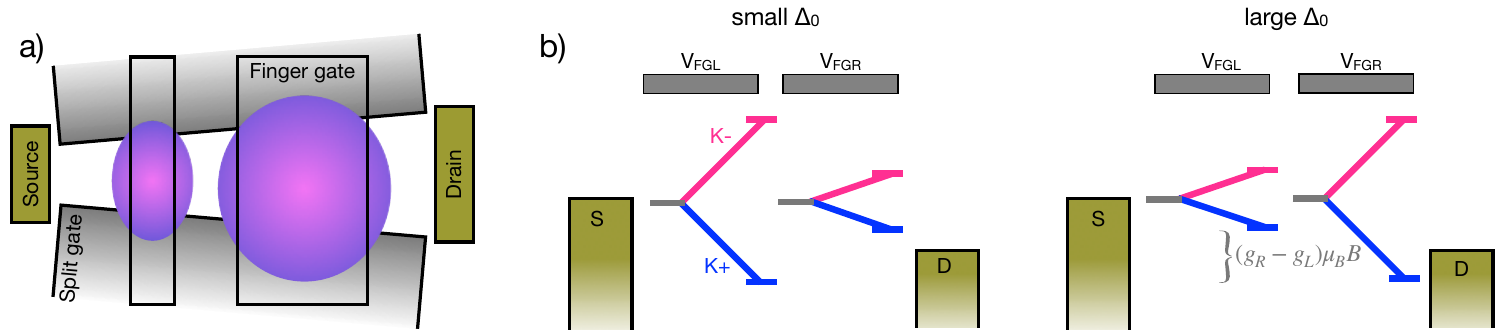}
  \caption{a) Double-dot setup of two electrostatically confined quantum dots defined by two split gates and two finger gates where the two dots have very different size and geometry. b) The different size and shape of the left and right quantum dots imply different valley g-factors and, hence, different valley splittings in a perpendicular magnetic field. Whether the g-factor decreases (for small gaps) or increases (for significant gaps) with dot size (cf. Figure \ref{fig:2}) may be detected in finite bias spectroscopy from the relative sign and ratio of the finger gate voltages on the left and right dot, $V_{FGL}$ and $V_{FGR}$, needed to re-establish tunnel transport when a weak magnetic field is applied.}
  \label{fig:C}
\end{figure}
 
 Our results demonstrate how the dot states and their properties can be manipulated by virtue of gate-tunable parameters. Such an understanding is required for tailoring specific, confined states, e.g., with the desired degeneracy, by quantum dot design.

For example, our findings of valley g-factor manipulation by quantum dot size may be exploited in multi-dot systems as the one sketched in \textbf{Figure \ref{fig:C} a)}, where different dot sizes and shapes imply different valley g-factors for the states in the left and the right dot.  Hence, valley g-factor modulation by quantum dot design allows for controllably induced dot-dependent state splittings in a constant perpendicular magnetic field, Figure \ref{fig:C} b). Similar to setups of semiconductor quantum dots with slanted magnetic fields \cite{katoGigahertzElectronSpin2003, pioro-ladriereElectricallyDrivenSingleelectron2008, pioro-ladriereSelectiveManipulationElectron2008}, device designs like the one in Figure \ref{fig:C} may enable, e.g., selective blocking and transport of specific states and single-site valley control without the need of an inhomogeneous B-field. We hence demonstrate how the simultaneous tuning of material properties and confined quantum states offers immense control over bilayer graphene quantum dots and holds great potential, e.g., for future valley-logic and valleytronics applications.



\medskip

\medskip
\textbf{Acknowledgements} \par 
We thank  Samuel Möller, Max Ruckriegel, Eike Icking, Katrin Hecker, Christoph Stampfer,  Vladimir Fal'ko, and Christoph Schulz for fruitful discussions. We acknowledge funding by the Deutsche Forschungsgemeinschaft (DFG, German Research Foundation) - SFB 1277 (Project No. 314695032).

\medskip
 \section*{Appendix: Data for extended parameters of the gap variation}
\begin{figure}[h!]
    \centering
    \includegraphics[width=\linewidth]{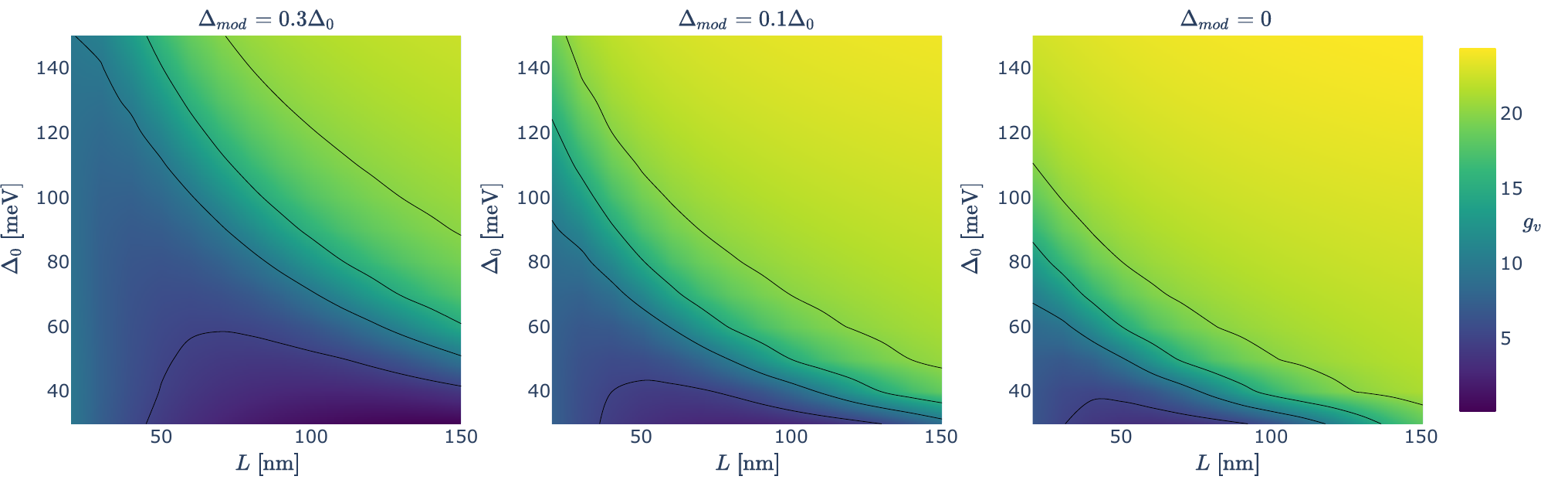}
    \caption{Lowest orbital dot states' valley g-factor over a range of the system parameters $L$ and $\Delta_0$ for different values of the gap profile parameter $\Delta_{mod}$. Here, $U_c = 5$ meV for all three cases.}
    \label{fig:add_data_Deltamod_diff}
\end{figure}

 In Figure \ref{fig:add_data_Deltamod_diff}, we demonstrate the stability of our results with respect to variations of the parameter $\Delta_{mod}$ in the gap profile in Equation \eqref{eqn:Potentials}. This parameter describes the modulation of the \BLG{} gap underneath the gates and can differ slightly between experiments, depending on the exact device architecture and screening. While varying $\Delta_{mod}$ leads to  shifts of the exact quantitative values of the boundary between the minivalley and valley center regime, see Figure \ref{fig:add_data_Deltamod_diff}, there are no qualitative changes. We expect our results and conclusions to hold as long as the potential is smooth and of the same symmetry.

%


\bibliographystyle{MSP}
\bibliography{QDgv.bib}

\end{document}